\input{aipcheck}

\documentclass[
    ,final            
  ]
  {aipproc}

\usepackage{amssymb,latexsym}
\usepackage{natbib}
\layoutstyle{6x9}

\newcommand{\heii}{He~{\rm II}}
\newcommand{\oiii}{[O~{\rm III}]}

\newcommand{\sii}{[S~{\rm II}]}

\newcommand{\Mpc}{~\ensuremath{ \rm Mpc \/}}
\newcommand{\pc}{~\ensuremath{ \rm pc \/}}
\newcommand{\kms}{~\ensuremath{ \rm km~s^{-1} \/}}

\newcommand{\Myr}{~\ensuremath{ \rm Myr \/}}
\newcommand{\ergl}{~\ensuremath{\rm erg~s^{-1}}}

\begin{document}

\title{Ultraluminous X-ray Sources and Their Nebulae}


\author{Pavel~Abolmasov}{
  address={Special Astrophysical Observatory, Russia}
}



\begin{abstract}
One of the interesting features of Ultraluminous X-ray sources is that
many of them are surrounded by luminous nebulae exhibiting diverse
observational properties. In different cases the nebulae are
photoionized or shock-powered. Generally, the two energy sources
appear to coexist. 
ULX bubble nebulae may be considered a new class of
shock-powered nebulae similar to upscaled versions of stellar wind
bubbles. Their expansion rates support constant
energy influx rather than single powerful events like Hypernova
explosions.
\end{abstract}

\maketitle


\section{Introduction}

We know of Ultraluminous X-ray sources (ULXs) since the late 80th. 
\cite{fabbiano} proposed that some of them could be young SNRs or recent
SNe mentioning that other objects require a different
explanation. 
In many cases spectral, spatial and timing
properties argue against the SNR explanation, as in the case of
NGC6946~X-1 \cite{RoCo}.

The object in NGC6946~\cite{schlegel_mf16} was not among the first ULXs discovered but 
probably the first to be identified in the optical range
\cite{bf_94}. For several years the object was considered an
exceptionally luminous SNR. 
 The second one was HoIX~X-1 identified with a complex nebula
MH9/10/11 \cite{miller}. 
Today, both remain unique: MF16 as the most
compact ULX nebula (ULXN) and MH9/10/11 due to abnormal physical properties and line
ratios in the second part of the nebula (MH11, see below). 

Direct optical identifications of ULXs are rare and challenge the
observational facilities of the largest telescopes. In several cases
optical or UV point sources have been detected but probably only in one
case \cite{liu_n5204} has a deep UV spectrum appropriate to determine
the spectral class been obtained.
The object (NGC5204~X-1) was classified as an early-B supergiant with some
peculiarities (see also discussion in \cite{mf16_pasj}). 

Optical nebulae are much easier to observe and may provide important information about
the central source, acting as calorimeters for the radiation and
the mechanical energy source (single explosion or wind/jet activity)
connected with the ULX.
For distances larger than $\sim 10 -- 20 \Mpc$ the nebulae become
difficult to study as well but the stellar population of the host cluster
of star-forming region is usually
appropriate for spectroscopy. For distant ULXs it is only possible to
study the host stellar population.

\section{ULX Nebulae}

Most of the achievements in studying both ULXs and their optical
counterparts were made in the last decade.
 First in Pakull and Mirioni \cite{PaMir} it
was realised that quite a number of sources may be identified with bubble
nebulae. Other ULXs are connected with photoionized nebulae that are
usually fainter in Balmer lines but require similar power $\sim 10^{39}\ergl$. 
In general both shock waves and photoionization by the central source
contribute.

It was realised that an enhanced \heii$\lambda$4686 line is common for
many ULXNe \cite{kaaret_hoii,kuntz} including even ULX bubbles
\cite{list,pakull_n1313}. 
The \heii\ emission is often proposed to be a wind or atmospheric
emission of the central object. Pakull et al. \cite{pakull_n1313} probably
detected radial velocity variations for the emission line in the
optical spectrum of NGC1313~X-2. This result may appear to be the
first detection of orbital motion for a ULX. 
Usually (see \cite{list} and \cite{mf16_pasj}) the \heii$\lambda$4686
emission is narrow ($FWHM \lesssim 200\kms$) but is emitted in
the vicinity of the X-ray source therefore the best explanation is a
compact HeIII-region. 

Several recent works on ULXNe such as Ramsey et al. \cite{ramsey} cover several
objects (up to 8). The number of more-or-less studied ULXNe is
slightly higher than this because clear examples are rare. Some of
ultraluminous X-ray sources definitely lack a conventional ULXN either
because of the different nature of the source or due to environmental
reasons (such as rarefied ISM or a crowded star-forming region).


\section{Host Clusters}

In many cases sources are found in young stellar clusters
\cite{ramsey,soria_n4559,abolswartz} implying that we are dealing
with young objects of stellar nature, probably with high-mass X-ray
binaries (HMXB). The youngest ages detected are about 4\Myr\
\cite{ramsey,abolswartz} but it is difficult to judge about the higher
age limit due to detectability problems. 
ULX nebulae residing in large star-forming regions are more difficult
to detect. In our work on NGC7331~X-1 ~\cite{abolswartz} that is
coincident with a massive young star cluster and HII-region we analyze the low excitation
forbidden line excess observed in the spectrum. It
may be partially ascribed to the existing interstallar shocks produced
by stellar winds and supernova remnants but additional
power is required several times higher than the contribution from the
several SNRs and wind bubbles predicted by stellar population
synthesis. The excess luminosities of the forbidden lines (both of high
and low excitation) are quite ordinary for ULX nebulae.

Another example of this kind is the star cluster $n5194-839$ from the
catalog of Larsen \cite{larsen} that coincides within several parsecs with an
X-ray source M51~X-7. The cluster was detected by Terashima et al. \cite{terashima2006}
but not identified with the Larsen object. According to Larsen
\cite{larsen}, the cluster is about $ 12\Myr$ old.
 Optical spectrophotometry \cite{list} reveals rather complicated stellar
 population spectrum with some excess in nebular lines.

\section{Kinematics}

The kinematics of ULXNe are known much worse than their line ratios and
physical conditions (that may usually be traced using diagnostic
lines). However it is a potentially important source of information
not affected (as line ratios are) by chemical composition.
Most expansion velocity estimates were made by
order of magnitude. Pakull and Mirioni \cite{PaMir} estimate the expansion rate of the
bubble of NGC1313~X-2 as $\sim 80\,\kms$ having long-slit
low-dispersion spectra. \cite{dunne} obtained
echelle spectra of MF16 and detected high-velocity wings at about
$200\,\kms$.
In the recent work by Ramsey et al. \cite{ramsey} echelle-spectroscopy of
MH9 is reported. Lines are found to be broadened by
$\sim 100\,\kms$. 
The importance of kinematical studies of ULXNe is in distinguishing
the two energy sources (shock waves and photoionization) that appear
to be acting simultaneously in most of the nebulae.

The best way to study more or less quiet kinematics of ULX bubbles is
by using scanning Fabry-P\'erot Interferometer (FPI) technique. Its advantage
against long-slit and echelle spectroscopy is in absence of slit
losses and additional information about the three-dimensional
properties of the objects.
In \cite{hoix_mois} we analyse our FPI data on MH9/10/11 (associated with HoIX~X-1)
with $\sim 30\,\kms$ spectral resolution in two forbidden
emission lines, \oiii$\lambda$5007 and \sii$\lambda$6717. We show that the
expansion of the nebula is anisotropic (velocities spanning the range
$30\div 70\,\kms$) and probably affected by density
gradients in the ISM. We also confirm the existence of the second
``bubble'' in the \oiii\ line (reported by \cite{pakull_beambags}) 
coincident with the faint nebula MH11. 
MH11 differs kinematically from MH9/10: velocity dispersion is
$\lesssim 10\,km\,s^{-1}$, and the line-of-sight velocity changes by
less than $20\,km\,s^{-1}$ within the nebula. Line ratios differ very
much, too: the luminosity of MH11 in \oiii$\lambda$5007 is about
ten times higher than in H$\beta$. We show that the
observed properties of MH11 may be explained by photoionization acting
in a rarefied medium with hydrogen density $n_H \sim 0.2 \div 0.3 \,\rm
cm^{-3}$. 

While we prepared the paper on MH9/10/11 an article by \cite{rosado}
appeared where quite a distant ULX was studied with a scanning FPI. Hence we
were not the first to apply this technique to ULXNe. However the authors only
detected two velocity components shifted by $\sim 60\kms$ with respect
to each
other. The question is open whether they detected an expanding shell
or two overlapping HII-regions with different line-of-sight velocities.

\section{The Nature of ULX Bubble Nebulae}

\begin{figure}
  \includegraphics[width=.9\textwidth]{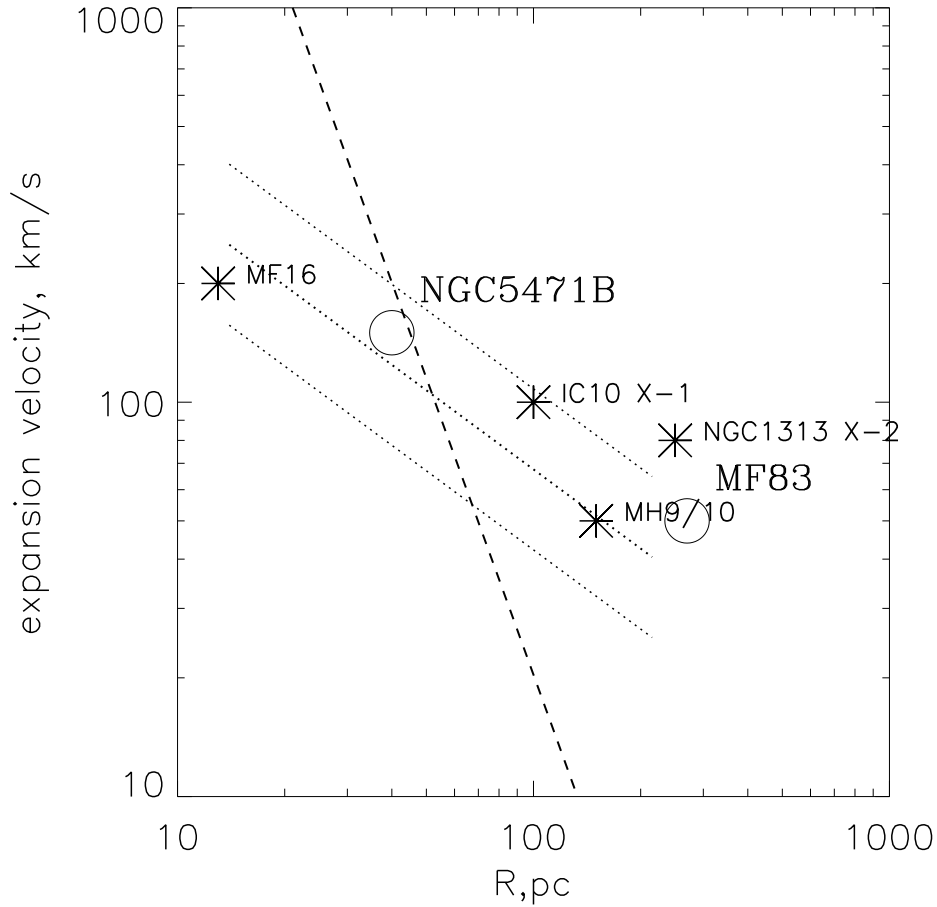}
  \caption{Radii and expansion velocities of ULX shells. Three dotted lines
  correspond to pressure-driven bubble solutions for $L=10^{39}\,erg\,s^{-1}$
  and ISM densities  $n = 1$, $10$ and $100\,cm^{-3}$. Dashed steep
  line corresponds to a radiative supernova solution with $E = 10^{52}\,erg$.
  Two HNR candidates in M101 are shown by circles.\label{fig:bubblevels}
}
\end{figure}

In Figure~\ref{fig:bubblevels} we show several ULX bubble nebulae with
measured expansion velocities. IC10~X-1 was added though it is not a
bona fide ULX but a more or less proven Wolf-Rayet HMXB with about-Eddington
X-ray luminosity \cite{ic10}. However the similarity between the bubble in
IC10 and ULX bubbles was noted by \cite{ic10} and appears to hold when
kinematical data are considered. Little velocity scatter and dependence
on the size indicate that
constant energy influx is a more likely solution for ULX bubbles.
The solution shown in Figure~\ref{fig:bubblevels} by dotted lines is
taken from \cite{castor} and is valid for adiabatic-stage and
pressure-driven bubbles:

\begin{equation}\label{E:predomR}
R = 110 n_0^{-1/5} L_{39}^{1/5} t_6^{3/5} \pc
\end{equation}
\noindent
\begin{equation}\label{E:predomV}
V = 64 n_0^{-1/5} L_{39}^{1/5} t_6^{-2/5} \kms
\end{equation}
\noindent
Here, $n_0$ is the hydrogen density of the unshocked
material. $L_{39}$ and $t_6$ are the kinematical luminosity in
$10^{39}\ergl$ and the  bubble age in millions of years.
The dependence between $V$ and $R$ is rather shallow in that case, $V
\propto R^{-2/3}$ and may be opposed to the solutions for SNRs.
$V\propto R^{-3/2}$  for Sedov solution and $V\propto R^{-5/2}$ for
the radiative stage \cite{lozinsk}. In the figure we show also the
pressure-driven snow-plough solution for a radiative SNR for comparison.  

Two well-known hypernova remnant (HNR) candidates in M101, NGC5471B \cite{chen} and MF83 \cite{mf83}, 
fall exactly in the
same region of the diagram. Due to that reason they are probable
candidates for ULXNe without the ULX phenomenon. In this scope it is
also reasonable to consider the bubble in IC10 not a remnant of a
powerful explosion \cite{lomo} but rather a relic ULX bubble. The
consistency between the observed properties of shell-like ULXNe may be
considered the strongest evidence for the nature of the enigmatic
sources. HMXB evolutionary scenarii predict that the ultimate stage of
evolution of an HMXB similar to SS433 is a Wolf-Rayet + Black Hole 
binary similar
to IC10~X-1 \cite{cherep95}. The characteristic evolutionary time is of
the order of 1\Myr\ consistent with the observed kinematical ages of
the oldest ULX bubbles. 

Supercritical accretor models for ULXs \cite{katz86,king2001,poutanen}
favour mild geometrical collimation of the emergent X-ray
radiation. Due to that reason the number of ULXNe is expected to be
higher than the number of ULXs -- off-axis ULXNe are expected to
exist. I would suggest to search for off-axis ULXNe among the
brightest SNRs having H$\alpha$ luminosity $\gtrsim 10^{37 -
38}\ergl$.

\section{Conclusions}

Most of the results on ULX nebulae and host clusters are related to a
limited number of spectacular sources like HoIX~X-1, HoII~X-1 and
NGC6946~X-1. Deeper studies of the well-known ULXNe reveal similarities
among these objects that appeared very diverse at first. For
example, photoionized regions of high excitation were discovered near
the ``classical'' bubble ULXN around HoIX~X-1 and appear to be created
by the same source. 

It seems that ULX bubbles are indeed bubbles blown by wind or jets rather than remnants of
powerful exposions. In addition to the high mechanical luminosity
photoionizing X-ray/EUV radiation of the central source is responsible
for some observed features of ULXNe. 

\begin{theacknowledgments}
I thank the organisers for supporting my participation in the
conference and the participants of the conference for numerous
fruitful discussions. My travel expences were partially covered
by RFBR 08-02-08085 travel grant. 
\end{theacknowledgments}



\bibliographystyle{aipproc}   

\bibliography{mybib}


\end{document}